\documentclass[12pt]{article}
\newlength{\mytopmargin}
\newlength{\myleftmargin}
\usepackage{amsmath,amsthm,amssymb,amsbsy,epsfig,graphicx,color,multicol,subfigure}
\usepackage{array,calc}
\usepackage[enableskew]{youngtab}
\usepackage{rotating}
\usepackage{young,epic}
\usepackage{a4wide,bm}
\usepackage{url}
\usepackage{hyperref}

\newtheorem{prop}{Proposition}

\usepackage{amsmath,amsfonts,amssymb}

\usepackage{graphicx}

\begin{document}
%

\title{Skew orthogonal polynomials for the real and quaternion real Ginibre ensembles and generalizations}
\author{Peter J. Forrester }
\date{}
\maketitle
\noindent
\thanks{\small Department of Mathematics and Statistics, 
The University of Melbourne,
Victoria 3010, Australia email:  p.forrester@ms.unimelb.edu.au 
}

\begin{abstract}
There are some distinguished ensembles of non-Hermitian random matrices for which the joint PDF can be
written down explicitly, is unchanged by rotations, and furthermore which have the property that the eigenvalues form a Pfaffian point process.
For these ensembles, in which the elements of the matrices are either real, or real quaternion, the kernel of
the Pfaffian is completely determined by certain skew orthogonal polynomials, which permit an expression in terms
of averages over the characteristic polynomial, and the characteristic polynomial multiplied by the trace. We use
Schur polynomial theory, knowledge of the value of a Schur polynomial averaged against real, and real quaternion
Gaussian matrices, and the Selberg integral to evaluate these averages.
\end{abstract}

\section{Introduction}

Let $X^{\rm r}$ be an $N \times N$ random matrix with elements independent and distributed as standard real Gaussians, and let $X^{\rm q}$ be
a $2N \times 2N$ random matrix with successive two-by-two blocks of the form
$$
\begin{bmatrix} z_{jk} & w_{jk} \\
- \bar{w}_{jk} & \bar{z}_{jk} \end{bmatrix}
$$
with each $z_{jk}$ and $w_{jk}$ independently chosen standard complex Gaussians. Equivalently, the joint probability density function (PDF) of
$X^{\rm r}$ is proportional to $e^{- {\rm Tr} X^{\rm r} (X^{\rm r})^T /2}$ and that of $X^{\rm q}$ is proportional to
$e^{- {\rm Tr} X^{\rm q} (X^{\rm q})^\dagger /2}$.  These are said to define the real and quaternion real Ginibre ensembles respectively \cite{Gi65}.
Notice that the distribution of the former is unchanged by either of the mappings $X^{\rm r} \mapsto X^{\rm r} O$ and $X^{\rm r} \mapsto O X^{\rm r} $, for
$O \in {\rm O}(N)$, while the latter is unchanged by  either of the mappings $X^{\rm q} \mapsto X^{\rm q} S$ and $X^{\rm q} \mapsto S X^{\rm q} $
for $S \in {\rm Sp}(2N)$. Real and quaternion real matrix ensembles possessing these properties will be denoted ${\rm RO}_N(g)$ and
 ${\rm QS}_{2N}(g)$ respectively, where $g$ is the underlying joint PDF of the matrices. Thus ${\rm RO}_N(e^{-{\rm Tr} \, X X^T/2})$ corresponds to the real
 Ginibre ensemble, and ${\rm QS}_{2N}(e^{-{\rm Tr} \, X X^\dagger/2})$ corresponds to the   quaternion real Ginibre ensemble.
 
 For both these ensembles, the corresponding eigenvalue distribution has the property that the corresponding $k$-point correlation functions
 are given by a $2k \times 2k$ Pfaffian, in which the elements depend on a certain kernel function independent of $k$. In the
 quaternion real case this was established in \cite{Ka98} and \cite[Exercises 15.9 Q.2]{Fo10}, and was earlier conjectured by Mehta \cite{Me91}.
 The understanding of this property for the real Ginibre ensemble came later, and is exhibited in the works \cite{FN07,Som07,BS09}.
 More recently ensembles of real and quaternion real matrices with joint PDFs proportional to
 \begin{equation}\label{14.1}
{ (\det G^\dagger G)^{\alpha_1} \over  \det ( \mathbb I + G^\dagger G)^{\alpha_2}}
\end{equation}
and
 \begin{equation}\label{14.2}
{ (\det G^\dagger G)^{\beta_1} \det (  \mathbb I - G^\dagger G)^{\beta_2}}, \qquad || G^\dagger G || < 1
\end{equation}
have been shown to also have the property that their eigenvalues form a Pfaffian point process; see \cite{FM09}, \cite{Ma12,FM13} in relation to the
former, and \cite{KSZ09}, \cite{Fi12} for the latter. We remark that the joint PDF (\ref{14.1}), for certain $\alpha_1, \alpha_2$ gives the
distribution of the matrix $G = A^{-1} B$, with $A$ and $B$ independent Ginibre matrices \cite{Kr06,FM09,Ma12}, while the joint 
PDF (\ref{14.2}), for certain $\beta_1$, $\beta_2$ gives the distribution of sub-blocks $G$ of unitary matrices with real, complex and real
quaternion elements \cite{ZS01,Fo06a, FK09,KSZ09,Fi12}. It is also
the case that these point processes can be naturally projected onto the sphere in the case of (\ref{14.1}) \cite{HKPV08,FM09} and
the anti-sphere in the case of (\ref{14.2}) \cite{FK09}, \cite{Ma11}, providing us with the names spherical and anti-spherical
ensembles respectively. 

Another construction of random matrices with joint PDFs (\ref{14.1}) and (\ref{14.2}) allows for $G$ to be rectangular. In the case of
(\ref{14.1}), let $A =X  X^\dagger$, where $X$ is an $M \times m_1$, $m_1 \ge M$ standard Gaussian matrix with real $(\beta = 1)$,
complex $(\beta = 2)$ or real quaternion $(\beta = 4)$ entries, and let $Y$ similarly denote such Gaussian matrices, but now with
$m_1$ replaced by $N$, $N \le M$. Then we know from \cite{GN99} and \cite[Exercises 3.6 q.3]{Fo10} that the rectangular random matrix
$G := A^{-1/2} Y$ has joint PDF given by (\ref{14.1}) with $\alpha_1 = 0$ and $\alpha_2 = (\beta/2)(m_1 + N)$. In relation to (\ref{14.2})
the rectangular matrix is literally a rectangular sub-block of the unitary matrices with real, complex and real quaternion elements
\cite{ZS01,Fo06a}; specifically, for a block $G$ of size $N \times M$ from a $K \times K$ unitary matrix (when the elements are real
quaternions, all these sizes should be doubled) the joint PDF (\ref{14.2}) results with $\beta_1=0$ and $\beta_2 = (\beta/2)(K-N-M+1-2/\beta)$.
Now form the square matrix $U (G G^\dagger)^{1/2}$, where $U$ is a random unitary matrix with the elements real, complex or
real quaternion as for $G$. This random matrix has the same joint distribution as $G$, but multiplied by the factor
$(\det G G^\dagger)^\alpha$ with $\alpha = (M-N)/2$ in the real case, and $\alpha = M - N$ in the complex and quaternion
real cases \cite{FBKSZ12}, and thus gives rise to (\ref{14.1}) and (\ref{14.2}) with non-zero values of $\alpha_1$ and $\beta_1$
\cite{FF11}, \cite{FM13} (see also Section 3.2 below; this class of square random matrices formed from rectangular random matrices is referred
to as an induced ensemble ).

An important point is that the 
generalized partition function for the joint eigenvalue PDF (in the real case this requires summing over the allowed number of
real eigenvalues---$0,2,\dots,N$ for $N$ even) can be expressed as a Pfaffian in which the matrix elements are a skew inner
product  $\langle \cdot, \cdot \rangle_{\rm s}$ of a polynomial basis of the monomials up to and including $x^{2N-1}$;
in the real case see \cite[Eqns.~8--10]{FN07}, \cite[Prop.~3.1]{FM09} and \cite[Eqns.~(3), (6)]{KSZ09}, while for the quaternion real case see
\cite{Ka98}, \cite{Ma12}, \cite{FM13}. The skew orthogonal inner product for all the real cases of the above joint PDFs is the same except for the
details of the one-body functions therein, and similarly for the quaternion real cases.
Moreover, the 
kernel function for the correlation functions is completely determined once the polynomials are chosen as $\{ Q_j(x) \}_{j=0,1,2,\dots}$,
each polynomial $Q_j(x)$ monic and of degree $j$, and having the skew orthogonality property
\begin{equation}\label{3.1}
\langle Q_{2j}, Q_{2k+1} \rangle_{\rm s} = r_j \delta_{j,k} \: \: (j \ge k) \qquad \langle Q_{2j}, Q_{2k} \rangle_{\rm s} =  \langle Q_{2j+1}, Q_{2k+1} \rangle_{\rm s} = 0.
\end{equation}

The common structure means that there is a common structure for the skew orthogonal polynomials. In the real case this was identified by
Akemann et al.~\cite[Eqns.~(4.6)--(4.7)]{AKP10} and reads
\begin{align}\label{4.1}
Q_{2n}^{\rm r}(z) & = \langle \det (z \mathbb I_{2n} - G) \rangle_{G \in {\rm RO}_{2n}} \nonumber \\
Q_{2n+1}^{\rm r}(z) & =  z Q_{2n}^{\rm r}(z) + \langle \det (z \mathbb I_{2n} - G) {\rm Tr} \, G \rangle_{G \in {\rm RO}_{2n}}
\end{align}
(in fact in general $Q_{2n+1}(z)$ is not unique --- any multiple of $Q_{2n}(z)$ can be added without affecting (\ref{3.1}).
Formally the same expressions hold in the quaternion real case, 
\begin{align}\label{3.2}
Q_{2n}^{\rm q}(z) & = \langle \det (z \mathbb I_{2n} - G) \rangle_{G \in {\rm QS}_{2n}} \nonumber \\
Q_{2n+1}^{\rm q}(z) & =  z Q_{2n}^{\rm q}(z) + \langle \det (z \mathbb I_{2n} - G) {\rm Tr} \, G \rangle_{G \in {\rm QS}_{2n}},
\end{align}
a fact known earlier than for the real case, and due to Kanzieper \cite{Ka98}.

It is our aim in this paper to show how to compute the matrix averages in (\ref{4.1}) and (\ref{3.2}) for the real and quaternion
Ginbre ensemble, its induced counterpart, as well as the real and quaternion real versions of the
spherical and anti-spherical
ensembles specified by the joint PDFs (\ref{14.1}) and (\ref{14.2}).

\section{Schur polynomials, matrix integrals and generalized Selberg integrals}
\setcounter{equation}{0}

Let $G$ be a square matrix. The theory of group characters tells us that there a special symmetric polynomial in the eigenvalues of $G$
--- the Schur polynomial --- indexed
by a partition $\kappa := (\kappa_1,\dots,\kappa_N)$, and denoted $s_\kappa(X)$.  One relatively simple property of the Schur polynomials
is the expansion
\begin{equation}\label{5.0}
\det ( z \mathbb I_N - G) = \sum_{l=0}^N z^{N - l} (-1)^l s_{1^l}(G),
\end{equation}
where $1^l$ denotes the partition with 1 repeated $l$ times and all other parts zero. This follows immediately once one is familiar with the fact
that $s_{1^l}(x) = e_l(x)$, where $x:=(x_1,\dots,x_N)$ and $e_l$ denotes the $l$-th elementary symmetric polynomial. The expansion
(\ref{5.0}) will be used to substitute for $ \det (z \mathbb I_{2n} - G) $ in (\ref{4.1}) and (\ref{3.2}).

This expansion in the second of the formulas in (\ref{4.1}) and (\ref{3.2}) leads to the product of a Schur polynomial and Tr$\, (X) =: e_1(X)$.
Such a product can be expanded in terms of Schur polynomials by appealing to the Pieri formula (see e.g.~\cite[Prop.~12.8.4]{Fo10})
\begin{equation}\label{5.3}
e_m(x) s_\kappa(x) = \sum_{\lambda \atop
\lambda/\kappa \, {\rm a \, vertical} \, m \, {\rm strip}}
s_{\lambda}(x).
\end{equation}
Here
$\lambda / \kappa$
consists of those boxes of the diagram of $\lambda$ which are not in $\kappa$ and this is said to be a 
vertical $m$-strip if
$\lambda/\kappa$ consists of $m$ boxes, all of which are in distinct
rows. Thus
\begin{equation}\label{8.1b}
s_{1^l}(G) {\rm Tr} \, (G) = s_{(2,1^{l-1})}(G) + s_{1^{l+1}}(G).
\end{equation}.

More sophisticated properties of the Schur polynomials are the evaluations of the matrix averages which result from these substitutions.
In fact for $X$ real and unchanged by $X \mapsto X O$ and $X  \mapsto O X$, 
$O \in {\rm O}(N)$ we have \cite[Eq.~(3.6)]{FR09} (see \cite{SK09} for the case $A = \mathbb I$)
\begin{equation}\label{9.1a}
\langle s_\mu(A X) \rangle_{X} =
\left \{ \begin{array}{ll} \displaystyle
{P_\kappa^{(2)}(A  A^T) \over
(P_\kappa^{(2)}(1^N))^2}\langle P_\kappa^{(2)}(X X^T) \rangle_X, &  \mu = 2 \kappa \\
0, & {\rm otherwise}, \end{array} \right.
\end{equation}
where the partition $2 \kappa$ is the partition obtained
by doubling each part of $\kappa$, and $P_\kappa^{(\alpha)}(X)$ is the one parameter generalization of the Schur polynomial
known as the symmetric Jack polynomial (the case $\alpha = 2$ as appears in (\ref{9.1a}) can equivalently be
defined as the zonal polynomials associated with the symmetric space $gl(N,\mathbb R)/ O(N)$); see e.g.~\cite[Ch.~12]{Fo10}.
Similarly  for  $X$  quaternion real and unchanged by $X  \mapsto X S$ and $X  \mapsto S X  $
$S \in {\rm Sp}(2N)$
we have \cite[Eq.~(3.8)]{FR09}
\begin{equation}\label{9.1c}
\langle s_\mu(A X) \rangle_{X} =
\left \{ \begin{array}{ll} \displaystyle
{P_\kappa^{(1/2)}(A A^T) \over                    
(P_\kappa^{(1/2)}(1^N))^2} \langle P_\kappa^{(1/2)}(X X^\dagger) \rangle_X, &  \mu = \kappa^2 \\
0, & {\rm otherwise}, \end{array} \right.
\end{equation}
where
$\kappa^2$ is the partition obtained by repeating each part of
$\kappa$ twice. 

The final main ingredient in our evaluations of the skew orthogonal polynomials are averages of Jack polynomials
with respect to Selberg type densities. To present these formulas, let ME${}_{\beta,N}(g(x))$ denote a matrix ensemble
with eigenvalue PDF proportional to
$$
\prod_{l=1}^N g(x_l) \prod_{1 \le j < k \le N} | x_k - x_j |^\beta.
$$
The Selberg weight corresponds to the case $g(x) = x^a (1 - x)^b$, $0 < x < 1$.
Furthermore, introduce the generalized Pochhammer symbol 
\begin{equation}\label{5.1x}
[u]_\kappa^{(\alpha)} = \prod_{j=1}^N {\Gamma(u - (j-1)/\alpha + \kappa_j) \over
\Gamma(u - (j-1)/\alpha) }.
\end{equation}
An integration formula conjectured by Macdonald and proved by Kaneko and Kadell (see \cite[Eq.~(12.153)]{Fo10} tells us that
\begin{equation}\label{5.1y}
\langle P_\kappa^{(\alpha)}(x_1,\dots,x_N) \rangle_{{\rm ME}_{2/\alpha,N}(x^{\lambda_1}(1-x)^{\lambda_2)}} =
 P^{(\alpha)}_{\kappa}(1^N) \frac{[\lambda_1 +(N-1)/\alpha +1]^{(\alpha)}_{\kappa} }
{[\lambda_1 + \lambda_2 +2(N-1)/\alpha +2]^{(\alpha)}_{\kappa}}.
\end{equation}
Also required is the limiting case ($\lambda_2 \to \infty$ and scaling of the integration variables)
\begin{equation}\label{5.1y1}
\langle P_\kappa^{(\alpha)}(x_1,\dots,x_N) \rangle_{{\rm ME}_{2/\alpha,N}(x^{\lambda_1} e^{-x})} =
 P^{(\alpha)}_\kappa(1^N)[\lambda_1 + (N - 1)/\alpha + 1]_\kappa^{(\alpha)},
\end{equation}
and the variant \cite{Wa05} (see also \cite{FS09}, \cite{FK07})
\begin{equation}\label{5.1y1a}
\langle P_\kappa^{(\alpha)}(x_1,\dots,x_N) \rangle_{{\rm ME}_{2/\alpha,N}(x^{\lambda_1} (1 + x)^{-\lambda_2}) } =
 P^{(\alpha)}_\kappa(1^N) {[\lambda_1 + (N - 1)/\alpha + 1]_\kappa^{(\alpha)} \over 
 (-1)^{|\kappa|} [-\lambda_2 + \lambda_1 + 2 + 2(N-1)/\alpha]_\kappa^{(\alpha)} }.
\end{equation}

\section{Skew orthogonal polynomials for the Ginibre ensembles}
\setcounter{equation}{0}
\subsection{The classical Ginibre ensembles}
Using the explicit form of the skew inner products, the skew orthogonal polynomials for the real, and quaternion real, Ginibre ensembles
have been computed in earlier literature.

\begin{prop}\label{P1}
We have \cite{FN07}, \cite[Prop.~5.10.4]{Fo10}
\begin{equation}\label{7.1}
Q_{2n}^{\rm r}(z) = z^{2n}, \qquad Q_{2n+1}^{\rm r}(z) = z^{2n+1} - 2n z^{2n-1}
\end{equation}
and \cite{Ka98}, \cite[Exercises 15.9 Q.2]{Fo10}
\begin{equation}\label{7.2}
Q_{2n+1}^{\rm q}(z) = z^{2n+1}, \qquad Q_{2n}^{\rm q}(z) =  2^n n! \sum_{l=0}^n {z^{2l} \over 2^l l!} .
\end{equation}
\end{prop}

Here we want to rederive these results by a direct evaluation of (\ref{4.1}) and (\ref{3.2}). Consider first the real Ginibre ensemble.
Substituting (\ref{5.0}) in the first formula of (\ref{4.1}) gives
\begin{equation}\label{7.3}
Q_{2n}^{\rm r}(z) = \sum_{l=0}^{2n} z^{2n - l} (-1)^l \langle s_{1^l}(G) \rangle_{G \in RO_{2n}(e^{-{\rm Tr} \, G G^T/2})}.
\end{equation}
According to (\ref{9.1a}) the only non-zero case of the average in (\ref{7.3}) is $l=0$, so giving $Q_{2n}^{\rm r}(z) = z^{2n}$ as stated in
(\ref{7.1}).

Substituting (\ref{5.0}) in the second  formula of (\ref{4.1}), then making use of (\ref{8.1b}) tells us that
\begin{align}\label{8.1}
& \langle \det ( z \mathbb I_N - G)  {\rm Tr} \, G \rangle_{G \in RO_{2n}(e^{-{\rm Tr} \, G G^T/2})} \nonumber \\
&
 = \sum_{l=0}^{2n} z^{2n - l} (-1)^l \Big ( \langle s_{(2, 1^{l-1})}(G) \rangle_{G \in RO_{2n}(e^{-{\rm Tr} \, G G^T/2})}
 +   \langle s_{1^{l+1}}(G) \rangle_{G \in RO_{2n}(e^{-{\rm Tr} \, G G^T/2})}  \Big ).
\end{align}
According to (\ref{9.1a}) the only non-zero case of the averages in (\ref{8.1}) is  $l=1$  in the first, and so we have
\begin{equation}\label{8.1a}
 \langle \det ( z \mathbb I_N - G)  {\rm Tr} \, G \rangle_{G \in RO_{2n}(e^{-{\rm Tr} \, G G^T/2})} 
 =  - {z^{2n-1} \over P_{1^1}^{(2)}(1^{2n})} \langle P_{1^1}^{(2)}(G G^T) \rangle_{G \in 
 RO_{2n} (e^{-{\rm Tr} \, G G^T/2})} .
\end{equation} 

Changing variables to $H = GG^T$ in the RHS of (\ref{8.1a}), which introduces the Jacobian factor $(\det H)^{-1/2}$
(see e.g.~\cite[Prop.~3.2.7]{Fo10}), then changing variables to the eigenvalues and eigenvectors of the positive definite
real symmetric matrix $H$  (see e.g.~\cite[Prop.~1.3.4]{Fo10}), shows that
\begin{align}\label{9.2}
 \langle P_{1^1}^{(2)}(G G^T) \rangle_{
 G \in RO_{2n} (e^{-{\rm Tr} \, G G^T/2})} & =  \langle P_{1^1}^{(2)}(x_1,\dots,x_{2n}) \rangle_{{\rm ME}_{1,2n}(x^{-1/2} e^{-x/2})} \nonumber \\
 & =  P_{1^1}^{(2)}(1^{2n}) \, (2n),
 \end{align}
where the final equality follows from (\ref{5.1y1}). Substituting this in (\ref{8.1a}), then substituting the result in the second formula of
(\ref{4.1}) shows that $Q_{2n+1}^{\rm r}(z) = z^{2n+1} - 2n z^{2n-1}$ as stated in
(\ref{7.1}). 

We now turn our attention to the quaternion real Ginibre ensemble. Substituting (\ref{5.0}) in the first formula of (\ref{3.2}) gives
$$
Q_{2n}^{\rm q}(z) = \sum_{l=0}^{2n} z^{2n-l} (-1)^l \langle s_{1^l}(G) \rangle_{G \in {\rm QS}_{2n}(e^{-{\rm Tr} \, G G^\dagger/2})} .
$$
Use of (\ref{9.1c}) tells us that the average is non-zero when $l$ is even only. Substituting this non-zero value gives
\begin{equation}\label{10.a}
Q_{2n}^{\rm q}(z) = \sum_{j=0}^n { z^{2 (n - j)} \over P_{1^j}^{(1/2)}(1^n) }  \langle P_{1^j}^{(1/2)}(G G^\dagger) \rangle_{G \in 
{\rm QS}_{2n} (e^{-{\rm Tr} \, G G^\dagger/2})} .
 \end{equation}
 Changing variables to $H = G G^\dagger$, which introduces the Jacobian factor $\det H$ (see \cite[Prop.~3.2.7]{Fo10}), then changing
 variables to the eigenvalues and eigenvectors of the positive definite self dual quaternion real matrix $H$ shows that
\begin{align}\label{10.2a} 
  \langle P_{1^j}^{(1/2)}(G G^\dagger) \rangle_{G \in 
{\rm QS}_{2n}  (e^{-{\rm Tr} \, G G^\dagger/2})} & =   \langle P_{1^j}^{(1/2)}(x_1,\dots,x_n) \rangle_{{\rm ME}_{4,n}(x e^{-x})} \nonumber \\
 & =  P_{1^j}^{(1/2)}(1^n) \, \prod_{l=1}^j(2(n-l+1)),
 \end{align}
where the final equality follows from (\ref{5.1y1}). Substituting in (\ref{10.a}), we find after simple manipulation that the second formula
of (\ref{7.2}) is reclaimed.

With regards to the first formula in (\ref{7.2}), (\ref{8.1}) with  RO${}_{2n}(e^{-{\rm Tr} \, G G^T/2})$ replaced by
$ {\rm QS}_{2n}(e^{-{\rm Tr} \, G G^\dagger/2})$, we see from (\ref{9.1c}) that only the average over $s_{1^{l+1}}(G)$, $l+1$ even,
gives a non-zero contribution. Thus we have
\begin{equation}\label{10.p}
 \langle \det ( z \mathbb I_N - G)  {\rm Tr} \, G \rangle_{G \in QS_{2n}(e^{-{\rm Tr} \, G G^\dagger /2})} 
 = - z \sum_{j=1}^{n} {z^{2(n - j)}    \over P_{1^j}(1^n) }  
  \langle P_{1^j}^{(1/2)}(G G^\dagger) \rangle_{G \in  {\rm QS_{2n}}
 (e^{-{\rm Tr} \, G G^\dagger/2})} .
 \end{equation} 
 Substituting (\ref{10.a}) and (\ref{10.p}) in the second formula of (\ref{3.2}) shows cancellation occurs, leaving us with
 $Q_{2n+1}^{\rm q}(z) = z^{2n+1}$ as stated in (\ref{7.2}).

\subsection{The induced Ginibre ensemble}
Let $G_{M \times N}^{\rm r}$, $(M > N)$ denote the $M \times N$ rectangular analogue of the construction of a real Ginibre matrix.
With $O \in O(N)$, it is known \cite{FBKSZ12} that the random matrix $G_I^{\rm r} := O ( (G_{M \times N}^{\rm r})^T G_{M \times N}^{\rm r})^{1/2}$
has a joint PDF proportional to
\begin{equation}\label{10.1}
(\det (G_I^{\rm r} )^T G_I^{\rm r})^{(M-N)/2} e^{- {\rm Tr} G^{\rm r}_I (G^{\rm r}_I)^T /2}.
 \end{equation} 
 A similar result holds true for quaternion real Gaussian matrices where, with $S \in {\rm Sp}(2N)$, one has that \cite{Ip13}
 the random matrix 
  $G_I^{\rm r} := O ( (G_{M \times N}^{\rm r})^T G_{M \times N}^{\rm r})^{1/2}$
has a joint PDF proportional to
\begin{equation}\label{10.2}
(\det (G_I^{\rm q} )^T G_I^{\rm q} )^{M - N} e^{- {\rm Tr} G^{\rm q}_I (G^{\rm q}_I)^\dagger /2}.
 \end{equation} 
 Thus in both cases the joint PDF for the classical Ginibre ensembles is generalized by the additional factors
 $(\det G G^\dagger)^\alpha$, for appropriate $\alpha$. We can use (\ref{4.1}) and (\ref{3.2}), now with $G$ chosen from the
 corresponding induced ensemble, to calculate the skew orthogonal polynomials.
 
 \begin{prop}\label{P2}
 With $\alpha = (M-N)/2$, we have for the induced real Ginibre ensemble
 \begin{equation}\label{11.2a}
 Q_{2n}^{\rm r}(z) = z^{2n}, \qquad Q_{2n+1}^{\rm r}(z) = z^{2n+1} - 2 (n+ \alpha) z^{2n-1},
 \end{equation}  
 and with $\alpha = M - n$, we have for the induced quaternion real Ginibre ensemble
  \begin{equation}\label{11.2b}
 Q_{2n+1}^{\rm q}(z) = z^{2n+1}, \qquad Q_{2n}^{\rm q}(z) = 2^n \Gamma(n+1+\alpha) \sum_{l=0}^n {z^{2l} \over 2^l \Gamma(l + 1 + \alpha)}.
 \end{equation}  
 \end{prop}
 
 \noindent
 Proof. \quad The derivation of the first result in (\ref{11.2a}) goes through exactly as for the corresponding result in (\ref{7.1}), which
 corresponds to the case $\alpha = 0$. For the second result in (\ref{11.2a}), we see that (\ref{8.1a}) again holds, provided we replace
 $G \in  RO_{2n}
 (e^{-{\rm Tr} \, G G^T/2})$ by $G \in  RO_{2n}
 ( (\det G G^T)^\alpha e^{-{\rm Tr} \, G G^T/2})$. Doing this implies that (\ref{9.2}) must be correspondingly modified,
 \begin{align}\label{9.2s}
 \langle P_{1^1}^{(2)}(G G^T) \rangle_{G \in RO_{2n}
  (\det G G^T)^\alpha 
 e^{-{\rm Tr} \, G G^T/2})} & =  \langle P_{1^1}^{(2)}(x_1,\dots,x_{2n}) \rangle_{{\rm ME}_{1,2n}(x^{-1/2+\alpha} e^{-x/2})} \nonumber \\
 & =  P_{1^1}^{(2)}(1^{2n})  \, (2(n+\alpha)),
 \end{align}
where the final equality follows from (\ref{5.1y1}). Substituting in the second formula of (\ref{3.2}) gives the second formula in (\ref{11.2a}).

Turning our attention now to the induced quaternion real ensemble, we see that (\ref{10.a}) again holds, but with
 $G \in  {\rm QS}_{2n}
 (e^{-{\rm Tr} \, G G^\dagger/2})$ by $G \in  {\rm QS}_{2n}
 ( (\det G G^\dagger)^\alpha e^{-{\rm Tr} \, G G^\dagger/2})$, and that
\begin{align}\label{10.2a} 
  \langle P_{1^j}^{(1/2)}(G G^\dagger) \rangle_{G \in {\rm QS}_{2n}
 ((\det G G^\dagger)^\alpha e^{-{\rm Tr} \, G G^\dagger/2})}  &=   \langle P_{1^j}^{(1/2)}(x_1,\dots,x_n) \rangle_{{\rm ME}_{4,n}(x^{1 + 2 \alpha} e^{-x})} \nonumber \\
 & =  P_{1^j}^{(1/2)}(1^n)  \,  \prod_{l=1}^j (2(n-l+1+\alpha)),
 \end{align}
where the final equality follows from (\ref{5.1y1}). Substituting in (\ref{10.a}) gives the second formula
in (\ref{11.2b}).

The derivation of the evaluation of $Q_{2n+1}^{\rm q}(z)$ proceeds in exactly the same way as done above for the evaluation of this polynomial in the
classical quaternion real Ginibre ensemble. Again there is cancellation, and we are left with the monomial evaluation $Q_{2n+1}^{\rm q}(z) = z^{2n+1}$.
\hfill$\square$

\section{Spherical and anti-spherical ensembles}
\setcounter{equation}{0}
The strategy used to derive the evaluations of the skew orthogonal polynomials for the Ginibre and induced Ginibre ensembles
can also be carried out for the real and quaternion real versions of the spherical and anti-spherical ensembles as specified by the joint PDFs
 (\ref{14.1}) and (\ref{14.2}). We consider first the case that $G$ is real.
 
 \begin{prop}
 For the joint PDF  (\ref{14.1}) with $G$ real we have
 \begin{equation}\label{16.1}
 Q_{2n}^{\rm r}(z) = z^{2n}, \qquad Q_{2n+1}^{\rm r}(z) = z^{2n+1} - z^{2n-1} {\alpha_1 + n \over \alpha_2 - (2n + \alpha_1 + 1/2)} ,
 \end{equation}  
 while for the  joint PDF  (\ref{14.2}) with $G$ real we have
\begin{equation}\label{16.2}
 Q_{2n}^{\rm r}(z) = z^{2n}, \qquad Q_{2n+1}^{\rm r}(z) = z^{2n+1} - z^{2n-1} {\beta_1 + n \over \beta_2 + (2n + \beta_1 + 1/2)} ,
 \end{equation}   
\end{prop}

\noindent
Proof. \quad The mechanism for the evaluation $Q_{2n}^{\rm r}(z) = z^{2n}$ in both cases is precisely the same as our derivation of the first result in
(\ref{7.1}).

Consider then the evaluation of $Q_{2n+1}^{\rm r}(z)$. The equations (\ref{8.1}) and (\ref{8.1a}) again hold but with
${\rm RO}_{2n} (e^{-{\rm Tr} \, G G^T/2})$ replaced by 
$$
{\rm RO}_{2n}(  (\det G^\dagger G)^{\alpha_1}/ \det ( \mathbb I + G^\dagger G)^{\alpha_2}) \quad
{\rm and} \quad  {\rm RO}_{2n} ((\det G^\dagger G)^{\beta_1} \det (  \mathbb I - G^\dagger G)^{\beta_2})
$$
for the spherical and anti-spherical ensembles respectively.
We similarly modify (\ref{9.2}) so that in relation to
(\ref{14.1}) it reads
\begin{align}\label{9.2u}
& \langle P_{1^1}^{(2)}(G G^T) \rangle_{
 G \in RO_{2n} (\det G^\dagger G)^{\alpha_1} / \det (  \mathbb I + G^\dagger G)^{\alpha_2})} 
 & =  \langle P_{1^1}^{(2)}(x_1,\dots,x_{2n}) \rangle_{{\rm ME}_{1,2n}(x^{-1/2+\alpha_1} (1 + x)^{-\alpha_2})} \nonumber \\
 & =  P_{1^1}^{(2)}(1^{2n}) {\alpha_1 + n \over \alpha_2 - (2n + \alpha_1 + 1/2)},
 \end{align}
 where to obtain the final equality use has been made of (\ref{5.1y1a}), and that in relation to (\ref{14.2}) it reads
 \begin{align}\label{9.2u}
& \langle P_{1^1}^{(1/2)}(G G^T) \rangle_{
 G \in RO_{2n} (\det G^\dagger G)^{\beta_1} \det (  \mathbb I - G^\dagger G)^{\beta_2}} 
 & =  \langle P_{1^1}^{(1/2)}(x_1,\dots,x_{2n}) \rangle_{{\rm ME}_{1,2n}(x^{-1/2+\beta_1} (1 - x)^{\beta_2})} \nonumber \\
 & =  P_{1^1}^{(1/2)}(1^{2n}) {\beta_1 + n \over \beta_2 + (2n + \beta_1 + 1/2)},
 \end{align}
 where here the final equality follows from (\ref{5.1y}). Substituting these formulas in the analogue of (\ref{8.1a}), then substituting the result in the second
 formula of (\ref{4.1}) gives the stated formulas for $Q_{2n+1}^{\rm r}(z)$.
 \hfill $\square$
 
 The real spherical ensemble, in the case $\alpha_1=0$, $\alpha_2 = N$
  was first studied in \cite{FM09}. However there a fractional linear transformation was used to map the
 domain from the real line and upper half plane to the unit circle and unit disk. A direct comparison with the results
 (\ref{16.1}) is therefore not possible; in fact the corresponding skew orthogonal polynomials were found
 to be monomials. However, in the recent Ph.D.~thesis \cite{Fi12} a study of the induced real spherical ensemble has
 been undertaken. Then there is no advantage in introducing a linear fractional transformation. By making use of working
 which has a number of steps in common to that presented above, it is found that in the case of the parameters
 $\alpha_1 = L/2$, $\alpha_2 = (m+L+N)/2$ \cite[Eqns.~(4.2.97), (4.2.98)]{Fi12}
 \begin{equation}
 Q_{2j}^{\rm r}(x) = x^{2j}, \qquad Q_{2j+1}^{\rm r}(x) = x^{2j+1} - {2j +L \over m -  2j - 1} x^{2j-1}.
 \end{equation}
 This is precisely what is given by (\ref{16.1}) with these choices of $\alpha_1, \alpha_2$, and $N \mapsto 2n$ in the latter.
 
 The anti-spherical ensemble, with $\beta_1 = 0$, $\beta_2 = (L-N-1)/2$ has been studied in 
 \cite{KSZ09,Fo10,Ma11}, and in \cite{Fo10x} the skew orthogonal polynomials were evaluated as
 \begin{equation}
 Q_{2j}^{\rm r}(x) = x^{2j}, \qquad Q_{2j+1}^{\rm r}(x) = x^{2j+1} - {2j \over L + 2j} x^{2j-1}.
 \end{equation}
 This is indeed the same result as is obtained by setting $\beta_1 = 0$,  $\beta_2 = (L - 2n - 1)/2$ in (\ref{16.2}). The corresponding
 induced ensemble has been studied in \cite{Fi12}, corresponding to the parameters $\beta_1 = L_1/2$ and $\beta_2 = (L_2 - N - 1)/2$.
 Indeed, substituting these values in (\ref{16.2}), the latter with $N \mapsto 2n$, we reclaim the explicit form of the skew orthogonal
 polynomials found therein \cite{Fi12}[(4.2.95),(4.2.96)].
 
 We conclude with the calculation of the skew orthogonal polynomials for the spherical and anti-spherical ensembles in the case that
 $G$ is quaternion real.
 
 \begin{prop}
 For the joint PDF  (\ref{14.1}) with $G$ quaternion real we have
 \begin{equation}\label{19.1}
  Q_{2n+1}^{\rm q}(z) = z^{2n+1}, \quad Q_{2n}^{\rm q}(z) = {\Gamma(\alpha_1 + n + 1) \over \Gamma(2n+{1 \over 2}+\alpha_1 - \alpha_2)}
  \sum_{j=0}^n (-1)^{n-j} {\Gamma(n+j+{1 \over 2} +\alpha_1-\alpha_2) \over \Gamma(j+1+\alpha_1)} z^{2j} ,
 \end{equation}  
 while for the joint  PDF  (\ref{14.2}) with $G$ quaternion real we have
 \begin{equation}\label{19.2}
  Q_{2n+1}^{\rm q}(z) = z^{2n+1}, \quad Q_{2n}^{\rm q}(z) = {\Gamma(\beta_1 + n + 1) \over \Gamma(2n+{1 \over 2}+\beta_1 + \beta_2)}
  \sum_{j=0}^n {\Gamma(n+j+{1 \over 2} +\beta_1+\beta_2) \over \Gamma(j+1+\beta_1)} z^{2j }.
 \end{equation}  
 \end{prop}
 
 \noindent
 Proof. \quad The mechanism for the result $ Q_{2n+1}^{\rm q}(z) = z^{2n+1}$ in both cases is precisely the same as that already
 revealed for the corresponding result in (\ref{7.2}).
 
 For the evaluation of $ Q_{2n}^{\rm q}(z)$, the expression (\ref{10.a}) again holds, upon replacing the quaternion real Ginibre
 ensemble ${\rm QS}_{2n}(e^{-{\rm Tr} \, G^\dagger G/2})$ by the ensemble of $2n \times 2n$ matrices with
 joint PDFs (\ref{14.1}) and (\ref{14.2}) as appropriate. Furthermore
\begin{align}\label{9.2u}
& \langle P_{1^j}^{(1/2)}(G G^T) \rangle_{
 G \in QS_{2n} (\det G^\dagger G)^{\alpha_1}/ \det (  \mathbb I + G^\dagger G)^{\alpha_2})} \nonumber \\
 & =  \langle P_{1^j}^{(1/2)}(x_1,\dots,x_n) \rangle_{{\rm ME}_{4,n}(x^{1+2 \alpha_1} (1 + x)^{- 2 \alpha_2})} \nonumber \\
 & =  P_{1^j}^{(1/2)}(1^n) (-1)^j {\Gamma(\alpha_1 + n + 1) \Gamma(2n-j + {1 \over 2} + \alpha_1 - \alpha_2) \over
 \Gamma(2n + {1 \over 2} + \alpha_1 - \alpha_2) \Gamma(n- j + 1 + \alpha_1) },
  \end{align}
 where to obtain the final equality use has been made of (\ref{5.1y1a}), and
 \begin{align}\label{9.2u}
& \langle P_{1^j}^{(1/2)}(G G^T) \rangle_{ 
 G \in QS_{2n} (\det G^\dagger G)^{\beta_1} \det (  \mathbb I - G^\dagger G)^{\beta_2})}  \nonumber \\
 & =  \langle P_{1^j}^{(1/2)}(x_1,\dots,x_n) \rangle_{{\rm ME}_{4,n}(x^{1+2 \beta_1} (1 - x)^{2 \beta_2})} \nonumber \\
 & =  P_{1^j}^{(1/2)}(1^n) {\Gamma(\beta_1 + n + 1) \Gamma(2n-j + {1 \over 2} + \beta_1 + \beta_2) \over
 \Gamma(2n + {1 \over 2} + \beta_1 + \beta_2) \Gamma(n- j + 1 + \beta_1) },
  \end{align} 
 with the final equality now following  from (\ref{5.1y}).   
 
 \section*{Acknowledgements}
 Conversations with Anthony Mays initiating this work are acknowledged, as is the 
financial support of  the Australian Research Council for the project `Characteristic polynomials in random matrix theory'.
Anthony Mays is also to be thanked for undertaking a careful reading.


\providecommand{\bysame}{\leavevmode\hbox to3em{\hrulefill}\thinspace}
\providecommand{\MR}{\relax\ifhmode\unskip\space\fi MR }
\providecommand{\MRhref}[2]{%
  \href{http://www.ams.org/mathscinet-getitem?mr=#1}{#2}
}
\providecommand{\href}[2]{#2}

\end{document}